\documentclass[final,10pt]{elsarticle}

\usepackage{graphicx}
\usepackage{amssymb}
\begin{document}
\begin{frontmatter}

%% Title, authors and addresses

%% use the tnoteref command within \title for footnotes;
%% use the tnotetext command for theassociated footnote;
%% use the fnref command within \author or \address for footnotes;
%% use the fntext command for theassociated footnote;
%% use the corref command within \author for corresponding author footnotes;
%% use the cortext command for theassociated footnote;
%% use the ead command for the email address,
%% and the form \ead[url] for the home page:
%% \title{Title\tnoteref{label1}}
%% \tnotetext[label1]{}
%% \author{Name\corref{cor1}\fnref{label2}}
%% \ead{email address}
%% \ead[url]{home page}
%% \fntext[label2]{}
%% \cortext[cor1]{}
%% \address{Address\fnref{label3}}
%% \fntext[label3]{}

\title{Three predictions on July 2012 Federal Elections in Mexico based on past
regularities}
\author[UAM-A]{H. Hern\'andez-Salda\~na}
\address[UAM-A]{\'Area de F\'isica Te\'orica y Materia Condensada, Dpto. de Ciencias B\'asicas,
Universidad Aut\'onoma Metropolitana at Azcapotzalco,
Av. San Pablo 180, 02200, M\'{e}xico D.F., Mexico.}
\ead{hhs@correo.azc.uam.mx}

\begin{abstract}
Electoral systems is subject of study for physicist and mathematicians
in last years given place to a new area: sociophysics. 
Based on previous works of the author on the Mexican electoral processes in 
the new millennium, he found three characteristics appearing along 
the 2000 and 2006 preliminary dataset offered by the  electoral 
authorities, named PREP: I) Error distributions are not Gaussian or Lorentzian, they 
are characterized for power laws at the center and asymmetric lobes 
at each side. II) The Partido Revolucionario Institucional (PRI) 
presented a change in the slope of the percentage of votes obtained when it
go beyond the $70\%$ of processed certificates; hence it have an improvement 
at the end 
of the electoral computation. III) The distribution of votes for the PRI
is a smooth function well described by Daisy model distributions of 
rank $r$ in all 
the analyzed cases, presidential and congressional elections in 2000, 2003 and
2006. If all these characteristics are proper of the Mexican reality
they should appear in the July 2012 process. Here I discuss some arguments
on why such a behaviors could appear in the present process.  
\end{abstract}

\begin{keyword}
%% keywords here, in the form: keyword \sep keyword
vote distribution \sep election  \sep opinion polls 
\sep error analysis \sep election forensics
%% PACS codes here, in the form: \PACS code \sep code
%\PACS 87.23.Ge \sep 89.75.-k
%% MSC codes here, in the form: \MSC code \sep code
%% or \MSC[2008] code \sep code (2000 is the default)

\end{keyword}

\end{frontmatter}

%   IMPORTANT: DO NOT ERASE
%   KEYS
% P1 PAN
% P2 PRI
% P3 PRD
% P4 ROBERTO CAMPA, NUEVA ALIANZA
% P5 PATRICIA MERCADO, ALTERNATIVA

\section{Justification}

Inspired by the courage of Borghesi \cite{Borghesi09a,Borghesi09b} on made predictions on some behaviors
in 2009 elections in France based in past regularities, I present my owns for the
July 2012 Mexican election. Unfortunately, the number of datasets onto 
these predictions 
are sustained are much more smaller than the French case, but the government's 
independent institution who is in charge of election organization is relatively 
new. Additionally, the predictions presented here are made on the Previous Electoral
results Program (PREP after its Spanish acronym of {\it Programa de Resultados Electorales 
Previos}) which have their present form departing in 2000. On how the PREP works and 
about its history see the official web page of IFE \cite{IFEweb,IFE_prep,PREP}

This systems offers the first electoral results as the information arrives to the
headquarters of authorities. The electoral authorities are grouped in the Instituto Federal Electoral, IFE. The Mexican elections are organized with polling stations distributed along the country and admits, by construction, around $750$ votes. The day of the
election, the polling stations start at $8:00$ hrs and close at $18:00$ hrs except
if there are voters remaining. The votes  are counted by citizen elected and trained 
by the IFE in the presence of the political parties representatives. Several data 
are hand written by the president of the polling station, the highest authority at 
the cabin, in the cabin certificate. These data are: 
Total number of
received ballots at the beginning of the electoral process (Br), number of
remaining (not used) ballots (Bs), number of voters (V), number of
deposited ballots per cabin (Bd) and the number of votes received for each
party/candidate,V$_i$. 
The certified is stamped outside the electoral package, which contain the physical votes.
The president carries the electoral package to the collecting stations at the 
district head, named CEDAT, {\it Centros de Acopio y Transmisi\'on de Datos}. The data are captured and send to the IFE's headquarters.

Hence, according to the IFE's we page\cite{IFE_prep}:
``El PREP NO cuenta votos, sino que captura y publica la informaci\'on asentada en las Actas de Escrutinio y C\'omputo por los ciudadanos que participan como funcionarios de casilla.(
The PREP DOES NOT count votes, but it captures and publishes the information seated in the certificates of scrutiny and calculation by the citizens who take part as civil servants of cabin)''. 
And, 
``Los resultados presentados por el PREP son preliminares, tienen un car\'acter informativo y no son definitivos, por tanto no tienen efectos jur\'idicos.
(The results presented by the PREP are preliminary, have an informative character and are not definitive, therefore they do not have juridical effects.)''\cite{IFE_prep}.
The definite results are counted during the Count by Distict, which will start in July
4th. However, PREP offers an opportunity to know and learn about the errors and the 
ways the democratic process is carried on. In July 2006, the avoided crossing
between the two main candidates percentage of votes at PREP's dataset caused the suspicion of a large fraud, mainly
by electronic ways. In this sense it is  of fundamental importance to understand it 
and to fulfill the expected reliability. 
On the process of July 2006 see \cite{Klesner} for a chronicle, and references \cite{Crespo, Pliego, Mochan, HernandezSaldanaE0, Aparicio06a} about the controversial results.Recently appeared a new book~\cite{Diaz-Polanco}. The bibliography is certainly 
incomplete but I focus mainly in peer reviewed works and sociphysics literature.

Notice that, for the present process,
``Por primera ocasi\'on, durante este Proceso Electoral Federal, el IFE pondr\'a a disposici\'on de la ciudadan\'ia, a trav\'es de Internet, la imagen digital de las AEC(Actas de Escrutinio y C\'omputo) de las m\'as de 143 mil casillas. (For the first occasion, during this Electoral Federal Process, the IFE will put at the disposal of the citizenship, across Internet, the digital image of the AEC ({\it voting certificates}) of more than 143 thousand cabins)''\cite{IFE_prep}.

\section{The Predictions}

The origin of the author interest  in electoral data was due to the suspicion of a Mega fraud in July 2006.
But, soon, I discovered the increasing interest of physicist and mathematicians in 
the field. But there is a lot theoretical work, many of them summarized in the 
report of Castellano \cite{CastellanoReport}, but not so much on real data.
Some of the references on actual data are \cite{Borghesi09a, Borghesi09b,Borghesi, Borghesi2010, Borghesi2012a,Borghesi2012b,CostaFilho2006,Castellano2007,HernandezSaldanaE1,Herrmann,HernandezSaldanaE0}. With the experience gained analyzing electoral data in Mexico, some regularities
appeared, and I hope that some of them are part of the Mexican electoral system
(I accept that the predictions could be wrong). 
The main regularities appeared during the studies of Mexican elections
\cite{HernandezSaldanaE0,HernandezSaldanaE1,HernandezSaldanaE2,HernandezSaldanaE3}
are presented in the next subsections.

\subsection{I) Errors could be epidemic in contemporary Mexican elections}

Suspicion of a large fraud in July 2006 force us to analyze the self consistency 
data contained in the dataset provided by IFE. In order to test the 
existence of anomalies in the presidential data we calculated the error of all the independent 
tests (summarized in Table \ref{tabla2}) and to contrast them with the results of
congressional elections of the same year and the presidential one in 
2000~\cite{HernandezSaldanaE0}.
The global behavior, characterized by a power law decay in the center 
and two asymmetric lobes at each side appears in {\it all} the cases, with 
small differences. This result is highly surprising since the presidential 
process was in complete suspicion of fraud, meanwhile the presidential 
process in 2000 was clear. Hence I present the following prediction:

\bigskip
{\sl Error distributions in self consistency tests of PREP dataset will be
described globally by a power law at the center and two asymmetric lobes 
at each side.}
\bigskip

\begin{table}
\begin{center}
\begin{tabular}{|c|c|}
\hline

B. received $-$ (B. not used $+$ Number of voters)& Br - (Bs + V) \\
B. received $-$ (B. not used $+$ B. deposited)& Br - (Bs + Bd) \\
B. received $-$ (B. not used $+$ Votes for each party)& Br - (Bs+$\sum_i$ V$_i$) \\
Number of voters $-$  B. deposited & V - Bd  \\
Number of voters $-$ Votes for each party& V - $\sum_i$ V$_i$ \\
B. deposited $-$ Votes for each party & Bd - $\sum_i$ V$_i$ \\
\hline

\end{tabular}
\protect\caption{\label{tabla2}Table of errors considered for a self consistency test of the PREP database in July 2000 and 2006. The prediction will run on the same 
distribution of errors for all the obtainable data during July 2012.
We abbreviated Ballots with B.. The variable $i$ stands for the number of votes obtained
for each party/candidate.
}
\end{center}
\end{table}

Each distribution of error is builded up by calculating the error defined 
in \ref{tabla2}
on each cabin and doing  the histogram of the values obtained, i.e., 
how many cabins have values of error equal to 0,1,2,$\cdots$. In other words,
we are seen for appearance and missing of votes. Notice that in 
the ideal case all the error distributions must be a Dirac delta function,
or no lack or excess of votes. 

For the figures see \cite{HernandezSaldanaE0} at the arxiv.

\subsection{II) The Partido Revolucionario Institucional (PRI) is a sprinter}

The PREP data is published in real time and reproduced by several information 
services. In the dataset the time of arrival to the capture center is recorded,
hence a graph of percentage of votes for each party against the time 
is possible, but it is much more easy to handle if we plot the percentage of 
processed certificates instead of time. In \cite{HernandezSaldanaE0} we 
found that the behaviour of PRI ruled the general vote decay of the other
political parties due to a change in the behaviour. Beyond $70\%$ of 
processed certificates PRI presents a revival, it changes its rate of grow.
This behavior is present in all elections in 2006 (Figure 1 and 2 in reference
\cite{HernandezSaldanaE0}) and the presidential 
of 2000 (Figure 9 at the same work). It is a well known fact that PRI receive 
a lot of votes in geographical regions with a high marginalization index(see 
for instance \cite{Pliego}), such 
a regions could have a slow process of cabins to the capture centers, explaining 
why PRI is a sprinter,i.e., it have a better performance at the end of the journey.
During all this years many of such marginal regions are governed by the PRI,
hence there is no reason to believe that the mechanism that gives the 
performance improvement had been missed.

With these arguments I propose the second prediction:

\bigskip
{\sl In the graph of percentage of vote against percentage of processed certificates 
the PRI will change its rate of grow around the $70\%$ of computed certificates. i.e.
this political party has a good sprint}
\bigskip

\subsection{III) The PRI have a smooth vote distribution}

During our analysis of electoral data, the smoothness of the vote distribution
for the PRI was matter of special atention. This distribution is the histogram of the number of polling 
stations with certain amount of votes. For comparison with probability 
distribution the amount of votes is ``unfolded'' or ``deconvoluted'' by using 
the average of number of votes (See reference \cite{HernandezSaldanaE2} for
an explanation but this procedure is standard in data treatment of complex 
quantum systems).
In \cite{HernandezSaldanaE1} I reported the smooth behavior of 
this party in federal elections 2000, 2003 and 2006 using the definitive
dataset of Count by District. A fitting with a model, named 
daisy\cite{HernandezSaldana}, of different
ranks was tested with success. This model depends only on the rank, $r$, as free parameter and is written for the nearest neighbour as:
\begin{equation}
P_{r} (x) = \frac{(r+1)^{r+1}}{\Gamma(r+1)} x^r \exp[-(r+1)x].
\end{equation}
With $r$ integer and $\Gamma(\cdot)$ the Gamma function.

However, the distribution of daisy models is a particular 
case of a more general distribution named gamma and characterized by 
two free parameters, $\alpha$ and $\theta$ \cite{GammaDistribution}: 
\begin{equation}
P_{\Gamma} (x) = \frac{x^{\alpha}}{\Gamma(\alpha) \theta ^{\alpha}} 
   \exp[-\frac{x}{\theta}].
\end{equation}
Here the free parameters are real numbers. Author's hope is
that this distribution fits better the distribution of vote, but 
the daisy models offers a more physical interpretations as explained 
in \cite{HernandezSaldanaE2,HernandezSaldanaE3}.
In this way, I present the third prediction:

\bigskip
{\sl The distribution of votes for PRI, in presidential and both chambers elections,
could be fitted by a smooth distribution, 
in general by a gamma distribution or for those distributions of Daisy models.}
\bigskip

Even when PRI lost presidential election in 2000 and 2004 there were few polling stations with a small number of votes for this political party, hence it is unlike (but not impossible) that such an event 
occurs in the present election when its presidential candidate had been at the 
head in all the polls. Notice that even when its candidate shall not win the 
election the  distribution of votes could show a polynomial grow.

\section{Acknowledgments and apologies}
This work was supported by PROMEP/SEP and projects from CBI-UAM-A. I  
thank to A. Kunold, J.L. Cardoso and E. V\'arguez for encouragement and 
support. All the errors are my fault. I apologize for the broken English 
and the lack of appropriate figures. I shall repair it for the next version.

\end{document}